\documentclass[aip,rsi,amsmath,amssymb,
reprint,%
]{revtex4-1}

\usepackage{color}
\usepackage{graphicx}
\usepackage{dcolumn}
\usepackage{bm}
\newcommand{\gpt}{\dot\gamma}

\begin{document}

\preprint{AIP/123-QED}

\title[Ultrafast ultrasonic imaging coupled to rheometry]{Ultrafast ultrasonic imaging coupled to rheometry: principle and illustration}

\author{Thomas Gallot}
\altaffiliation[Now at ]{Department of Earth, Atmospheric and Planetary Sciences, Massachusetts Institute of Technology, 77 Massachusetts Avenue, Cambridge, MA 02139-4307, USA.}
\author{Christophe Perge}
\author{Vincent Grenard}
\author{Marc-Antoine Fardin}
\author{Nicolas Taberlet}
\author{S\'ebastien Manneville}
\altaffiliation[Also at ]{Institut Universitaire de France}
\email{sebastien.manneville@ens-lyon.fr}
\affiliation{Universit\'e de Lyon, Laboratoire de Physique, \'Ecole Normale Sup\'erieure de
Lyon, CNRS UMR 5672, 46 All\'ee d'Italie, 69364 Lyon cedex 07, France.}

\date{\today}

\begin{abstract}
We describe a technique coupling standard rheology and ultrasonic imaging with promising applications to characterization of soft materials under shear. Plane wave imaging using an ultrafast scanner allows to follow the local dynamics of fluids sheared between two concentric cylinders with frame rates as high as 10,000 images per second, while simultaneously monitoring the shear rate, shear stress, and viscosity as a function of time. The capacities of this ``rheo-ultrasound'' instrument are illustrated on two examples: (i) the classical case of the Taylor-Couette instability in a simple viscous fluid and (ii) the unstable shear-banded flow of a non-Newtonian wormlike micellar solution.
\end{abstract}

\pacs{07.64.+z,43.58.+z,83.85.-c}
\maketitle

\section{\label{s.intro}Introduction}

Soft matter is a flourishing field that involves an ever increasing number of characterization tools. Among them, techniques that allow one to follow the dynamics of complex fluids under shear are required to better understand the coupling between the shear flow and the material microstructure, i.e. its organization at the supramolecular scale \cite{Larson:1999}. Indeed, non-Hookean and/or non-Newtonian features in soft materials typically arise from the modification of the microstructure due to deformation and flow. In turn, microstructural changes lead to modifications of the fluid viscosity and hence of the flow field. For instance, the alignment and elongation of polymer molecules by shear are well known to lead to a reduction of the viscosity, known as {\it shear thinning}, in polymeric solutions. In some cases, due to this feedback mechanism between flow and microstructure, a homogeneous shear flow may become unstable and give way to a heterogeneous flow constituted of several bands of different local viscosities. Such a phenomenon, referred to as {\it shear banding} in the literature, turns out to be widespread among very different classes of materials, ranging from colloidal gels, emulsions, and pastes to self-assembled surfactant and polymer systems \cite{Olmsted:2008,Manneville:2008,Schall:2010}. Yet, shear banding is just one among several examples of complex flows found in soft materials, which may also display fractures in the bulk or apparent slip at the bounding walls \cite{Magnin:1990, Pignon:1996,Grondin:2008,Seth:2008,Lettinga:2009,Tabuteau:2009,Buscall:2010,Seth:2012}. The complete understanding of the physics of such dynamical and heterogeneous phenomena calls for space- and time-resolved investigations of the flow properties in well controlled geometries.

Rheometry is the most widespread tool to characterize the flow properties of a soft material in well-controlled geometries \cite{Barnes:1989,Macosko:1994}. Shear is generally applied by a rotating tool attached to a measuring device involving an air or a magnetic bearing. The most common shear cell geometries are (i) the cone-and-plate, where the sample is confined between a rotating cone and a fixed plate, (ii) the plate-and-plate, where the sample is confined between a rotating plate and a fixed plate, and (iii) the Taylor-Couette geometry, where the sample is confined between a fixed outer cylindrical cup and an inner rotating cylindrical bob. Rheological data captured by a rheometer include the shear strain $\gamma$, the shear rate $\gpt={\rm d}\gamma/{\rm d}t$ and the shear stress $\sigma$ experienced by the material, which are respectively computed from the position of the rotating tool, its rotation speed and the torque exerted on the tool. Standard rheological protocols consist in applying either a constant $\gpt$ ($\sigma$ resp.) and to measure the corresponding $\sigma$ ($\gpt$ resp.), from which the apparent shear viscosity $\eta=\sigma/\gpt$ is deduced as a function of time, or to impose an oscillating $\gamma$ or $\sigma$, from which the viscoelastic moduli $G'$ and $G''$, known as the elastic and viscous moduli, are respectively computed as the in-phase and quadrature-phase component of the measured response. For all cell geometries the proportionality factors that link the rotation speed and the torque to the shear rate and to the shear stress are calculated based on various assumptions on the flow and on the material. In particular, both the flow and the material are assumed to remain homogeneous. The reader is referred to standard rheology textbooks for more details \cite{Barnes:1989,Macosko:1994}. Our main point is that rheological measurements are essentially blind to local microstructural and/or flow heterogeneities since it provides only spatially averaged observables. Therefore complementary tools are needed to investigate complex flows such as the shear banding or fracture flows mentioned above.

In the last two decades a number of combined techniques have been introduced to go beyond standard rheology and get simultaneous local information on the microstructure and on the flow field. These have been recently reviewed in \cite{Manneville:2008}. Let us mention local structural characterization under shear through light\cite{Salmon:2003b}, neutron\cite{Liberatore:2006} and X-ray\cite{Welch:2002} scattering or through birefringence measurements\cite{Cappelaere:1997,Lerouge:2004}, as well as local velocity measurements through optical particle tracking\cite{Hu:2005,Miller:2007,Boukany:2010}, ultrasonic velocimetry\cite{Manneville:2004a}, or magnetic resonance imaging \cite{Callaghan:2008}. Among them the ultrasonic speckle velocimetry technique (USV) developed by one of us led to a number of important findings on shear banding flows thanks to its fast temporal resolution \cite{Becu:2004,Becu:2007}. The aim of the present paper is to extend our previous one-dimensional technique \cite{Manneville:2004a} to a combination of rheometry and ultrafast ultrasonic imaging leading to a two-dimensional time-resolved characterization of the shear flow of soft materials. In Sec.~\ref{s.setup} we describe the experimental setup used to acquire simultaneously global rheological data and ultrasonic images of the sheared material with emphasis on the technical features of our ultrasonic technique and on the processing of the acoustic data. In Sec.~\ref{s.newtonian} the instrument is first calibrated at low shear rates on a Newtonian fluid, a dilute suspension of hollow glass spheres. We further show that it can be used to image the toroidal vortices that appear above the critical shear rate corresponding to the onset of the classical Taylor-Couette instability. Finally Sec.~\ref{s.wlm} illustrates the capabilities of the technique in the case of a semidilute surfactant solution known to present both shear-banding and viscoelastic instabilities~\cite{Fardin:2012d}.

\section{Experimental set up}
\label{s.setup}
\begin{figure}
\includegraphics[width=7cm]{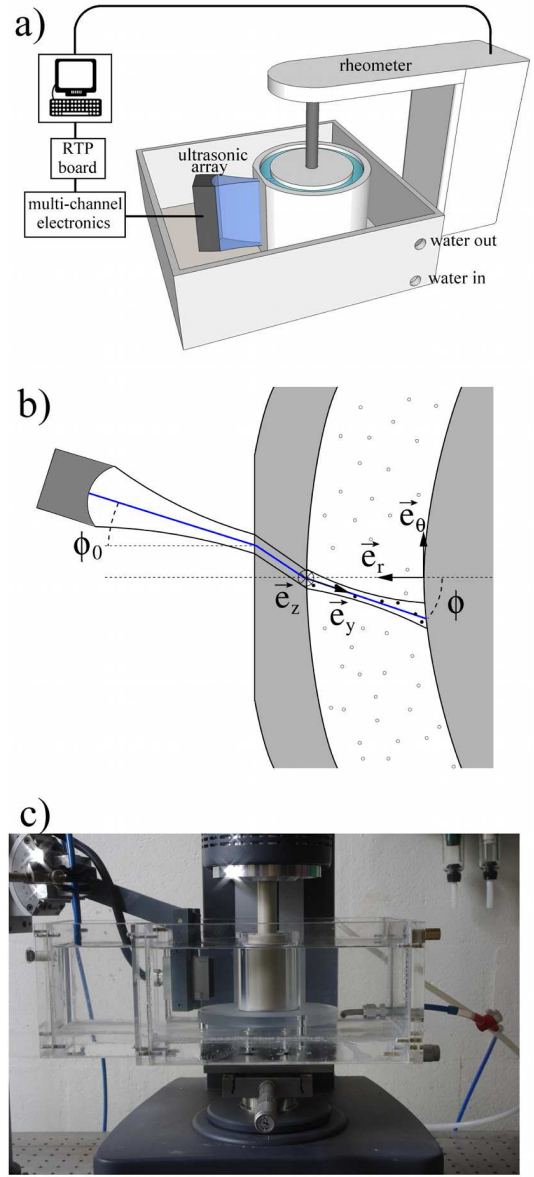}
\caption{\label{f.setup} Sketch and picture of the experimental setup. (a)~Three-dimensional general view. (b)~Top view of the gap of the Couette cell together with the path of the acoustic beam and the various axes and angles defined in the text. (c)~Picture showing the water tank with the ultrasonic transducer facing a smooth, transparent Taylor-Couette geometry with dimensions ($R_1=23$~mm and $R_2=25$~mm) smaller than that used in the text.}
\end{figure}

Our apparatus consists of a custom-made ultrasonic scanner coupled to a commercial rheometer. It is derived from our previous one-dimensional ultrasonic velocimetry technique \cite{Manneville:2004a}. Figure~\ref{f.setup}(a) shows the general design of the various instruments involved in our technique. Below we briefly discuss the rheometer and the rheological cell, which are both standard in the field of complex fluid characterization. Then we focus in detail on the ultrasonic imaging system.

\subsection{Rheological measurements}
\label{s.rheo}
The rheometer is a state-of-the-art stress-imposed rheometer with a magnetic air bearing (TA Instruments ARG2). It is equipped with a custom Taylor-Couette cell of height 50~mm. The radius of the inner rotating bob made of PEEK is $R_1=48$~mm while the fixed cup made of PMMA has a radius $R_2=50$~mm, yielding a gap $e=R_2-R_1=2$~mm. In a cylindrical geometry the shear stress $\sigma$ is not perfectly constant across the gap: it rather decreases as $1/(R_1+r)^2$, where $r$ is the distance to the inner bob. In our case this leads to a decrease of $\sigma$ by $\delta\sigma/\sigma\simeq 2e/R_1\simeq 8$~\% from the (inner) bob to the (outer) cup, which can be reasonably neglected so that the stress field can be taken as quasi-homogeneous. Also note that the bottom of the inner bob is machined as a cone of angle 2.3$^\circ$ and truncated at 50~$\mu$m from its tip, so that the shear rate below the rotating bob matches the average shear rate across the gap. This concentric cylinder cell terminated by a cone-and-plate type of geometry at the bottom is referred to as a Mooney-Couette cell in the rheology literature \cite{Barnes:1989,Macosko:1994}. Moreover it is well-known that the physical-chemical properties of the cell walls may affect drastically both global rheological properties and local flow behaviors \cite{Buscall:1993,Seth:2008,Gibaud:2008,Buscall:2010}. In particular wall roughness can be used to avoid slippage of the sample on the rotating bob and/or on the fixed cup. Here a ``smooth'' (polished) bob is used together with a ``rough'' (sand-blasted) cup. 

The cell is immersed in a large water tank (total volume of 2.2~L) connected to a water bath (Huber Ministat 230-CC-NR) that keeps the temperature constant to within 0.1$^\circ$C [see Fig.~\ref{f.setup}(a,c)]. This water tank is also used to provide acoustic coupling with a fair impedance matching between the sample contained in the cell and the ultrasonic transducer array (see below). The temperature at which the present experiments are conducted is 20.2$^\circ$C everywhere except in Sect.~\ref{s.wlm}. Finally the thickness of the cup is 5~mm everywhere except for a rectangular region milled to provide a minimal thickness of 1~mm and thus prevent a too strong attenuation of the ultrasound when travelling across the cup.

When illustrating the technique in Secs.~\ref{s.newtonian} and \ref{s.wlm} we shall only use the rheometer in the ``shear startup'' mode whereby a given steady shear rate is imposed starting from rest at time $t=0$ and the subsequent stress response $\sigma(t)$ is monitored. However it is clear that the technique can be adapted to steady shear from any initial shear rate other than zero, to ``step strain'' experiments where a given strain is imposed at $t=0$, to ``creep and recovery'' experiments where the shear stress is controlled rather than the shear rate, or even to oscillatory shear experiments. In all the experiments discussed below the target shear rate is reached by our stress-imposed rheometer using a feedback loop on the bob rotation speed. Note that, although the characteristic time of the feedback loop is of the order of 10~ms, it may take a much longer time before the target shear rate is actually reached by the rheometer due to the coupling between the geometry inertia and the fluid viscoelasticity. With the Couette geometry described above, we found this time to be of the order of 0.5~s in water [see Figs.~\ref{f.laminar}(b) and \ref{f.unstable}(b)] and about 3~s in viscoelastic wormlike micelles [see Figs.~\ref{f.micelles}(b)].

\subsection{Ultrasonic scanner and probe}
\label{s.scanner}
Our ultrasonic scanner is built upon a custom-made multichannel electronics designed for phased-array applications by Lecoeur Electronique. It consists of 128 independent channels, each made of a transmitter and a receiver. The transmitter can be either a ``spike'' transmitter that emits short pulses of tunable voltage (from 10 to 230~V) and width (from 25~ns to 3~$\mu$s, fall time less than 10~ns) and whose sequence in time is programmable with a maximum pulse repetition frequency ($f_{\rm PRF}$) of 20~kHz, or a fully programmable analog transmitter that generates power-amplified arbitrary waveforms with a bandwidth of 10~MHz and a maximum peak-to-peak amplitude of 100~V. In the latter case a 4 Mb memory is used to store the waveform which is sampled at 80~MHz and converted by a 12-bit linear digital-to-analog converter. Analog transmitters allow for transmitting high acoustic power at central frequencies up to 10~MHz while pulsed emission is used whenever a larger emitting bandwidth is needed, e.g. for ultrasonic imaging at frequencies larger than 10~MHz. In the present paper only pulsed emission is used to drive a high-frequency transducer array as described below.

The receiver is composed of a 12-bit linear analog-to-digital converter with a sampling frequency $f_s=160$~MHz and a bandwidth of 1 to 30~MHz, an ultra-low noise amplifier with a maximum gain of 80~dB, and a SDRAM memory of 32~Msamples. The input impedance is 50~$\Omega$. A real-time processor board is also featured, which allows for a real-time processing of ultrasonic data (not used in the present work) and for faster data transfer to the computer. The ultrasonic scanner is programmed under Matlab and the acquired data are stored on the PC hard drive for post-processing as described below in Sec.~\ref{s.process}.

The ultrasonic probe used in the present work is a custom-made linear array of 128 piezoelectric transducers designed by Imasonic. The transducers are 200~$\mu$m wide and spaced by 50~$\mu$m, resulting in a total active length of 32~mm. They work at a central frequency $f=15$~MHz with a bandwidth at -6~dB of about 8~MHz. The wavelength in water is thus $\lambda=c/f\simeq 100~\mu$m, with $c=1480$~ms$^{-1}$ the sound speed in water at 20$^\circ$C.\cite{Pierce:1994} In the elevation direction, i.e. in the direction perpendicular to the imaged plane, the transducers have an aperture $D=10$~mm with a cylindrical shape of radius 30~mm (see sketches in Fig.~\ref{f.setup}). This leads to a focusing in the elevation direction at a distance $F=30$~mm from the probe and over a focal spot whose length $l_f$ and width $w_f$ at -6~dB can be estimated respectively as $l_f\simeq8\lambda F^2/D^2\simeq 7$~mm and $w_f\simeq\lambda F/D\simeq 300~\mu$m.

In the following the region of interest, namely the gap of the Couette cell, will be centered in the focal spot and will be thus considered as a thin rectangular slice of height 32~mm, length $e=2$~mm, and thickness $300~\mu$m. In order to detect a non-zero displacement of the sheared material, the transducer array is set at an angle $\phi_0\simeq 5^\circ$ relative to the normal of the outer cup. After refraction in the cup, the actual angle of incidence within the sample, noted $\phi$, is close to but different from $\phi_0$ as shown in Fig.~\ref{f.setup}(b). This parameter will be precisely determined through the calibration procedure described in Sect.~\ref{s.calib}.

\subsection{Ultrasonic data processing}
\label{s.process}

\begin{figure*}
\includegraphics[width=12cm]{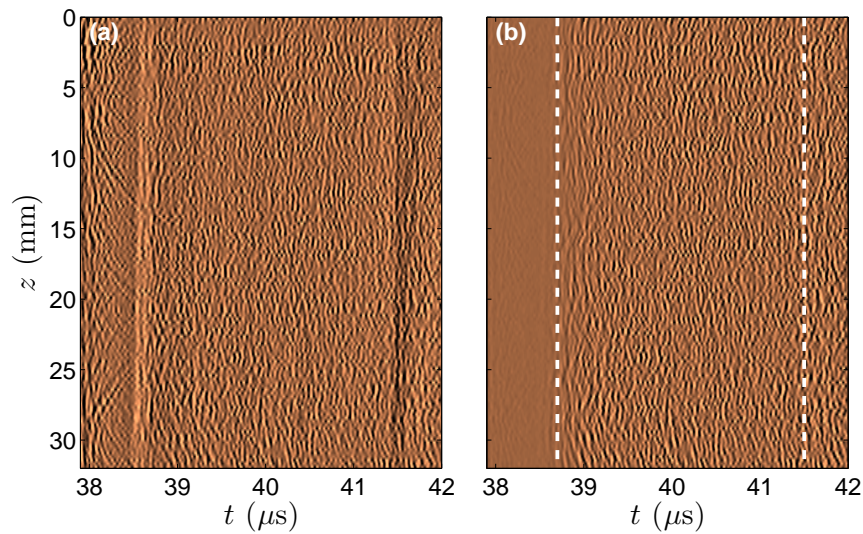}
\caption{\label{f.bscan} (a) Raw speckle signal $s_i(t,z)$ recorded after a single plane wave emission as a function of time $t$ and transducer position $z$ along the array. (b) Corrected speckle signal $\tilde{s}_i(t,z)$ after removal of the fixed echoes in $s_i(t,z)$ (see text). The dashed lines at $t=38.7$ and 41.5~$\mu$s indicate the limits of the gap as inferred from the calibration procedure described in Sect.~\ref{s.calib}. The signals are coded using linear color levels. Experiment performed in a Newtonian suspension of hollow glass spheres at 1~wt.~\% in water and sheared at $\dot\gamma=10$~s$^{-1}$ (see also supplementary movie 1)~\cite{Remark:moviesRSI}.}
\end{figure*}

\subsubsection{Plane wave imaging}
\label{s.plane}

Standard ultrasonic imaging schemes use a series of emissions focused at different locations in the sample and from which a single image is reconstructed. Doppler imaging can then be performed to measure one component of the flow velocity \cite{Jensen:1996}. In more recent ultrasonic velocimetry approaches successive images are stored and cross-correlated as in optical Particle Imaging Velocimetry (PIV) to infer the two-dimensional displacement field between two images. In general this ``EchoPIV'' technique is based on a standard scanner and yields a maximum frame rate of about 200~Hz which is enough for biomedical imaging and diagnosis \cite{Kim:2004} but too slow to follow transient phenomena or fast dynamics in soft matter. High-resolution imaging systems with working frequencies as high as 30~MHz now allow for ``micro EchoPIV'': Poiseuille flows in stenotic vessel-mimic phantoms of diameter 0.6~mm and blood flow in small animals have been imaged with a spatial resolution of 60~$\mu$m and frame rates up to 100~Hz \cite{Qian:2010}.

Our imaging technique relies on the use of ``plane wave imaging'' as introduced by Sandrin {\it et al} \cite{Sandrin:1999} in order to increase the temporal resolution of standard ultrasonic imaging by a factor of 10 to 100. A single emission of a plane wave is generated by firing all the transducers simultaneously. The signal sent to the transducers is constituted of a single pulse of amplitude 100~V. The echoes backscattered by the sample are amplified with a gain of 60~dB, recorded, and stored on the on-board memories of the various electronic channels. The pulsed plane wave emission is repeated $N$ times with a repetition frequency $f_{\rm PRF}$, which constitutes one ``sequence.'' Identical sequences can be repeated up to $N_{\rm seq}=512$ times. The number of pulses $N$ per sequence is limited to 8,192 emissions and by the total memory of 32 Msamples available per channel. Once the full series of imaging sequences is completed, the ultrasonic data are downloaded to the PC for post-processing. 

Such plane wave imaging has been used to follow shear waves into the human body and infer useful information on local elastic properties, a technique known as ``transient elastography,'' which was recently taken to real-time thanks to graphical processing units \cite{Montaldo:2009}. The pulse repetition frequency $f_{\rm PRF}$, hence the frame rate, is only limited by the total time-of-flight from the ultrasonic probe to the inner bob and back to the probe. In the present experimental setup the total travel distance is about 60~mm, which corresponds to a time-of-flight of about 40~$\mu$s and to a maximum frame rate of 25~kHz.

In our case, the aim is to follow the deformation and flow of soft materials. The backscattered ultrasonic signal either arises from the fluid microstructure itself or is obtained artificially by seeding the sample with microspheres that play the role of acoustic contrast agent. In both cases we call this signal an ``ultrasonic speckle'' and the corresponding velocimetry technique ``ultrasonic speckle velocimetry'' (USV). Note that a seeding concentration of a few volume percent generally provides a good signal-to-noise ratio and does not constitute any significant limitation of USV when compared to EchoPIV \cite{Zheng:2006}. Still plane wave imaging leads to a strong loss of lateral resolution, which is minimized in the following by a reception beam-forming, so that speckle tracking in USV is usually performed only on the axial component of the velocity.

\subsubsection{From raw speckle data to ultrasonic images}
\label{s.image}

\paragraph{Raw speckle signal. -} The ultrasonic speckle backscattered from the sample to the transducer array results from the interference of all scattering events of the incident plane wave by the scatterers across the insonified slice (32~mm $\times$ 2~mm $\times$ 300~$\mu$m). If one can neglect echoes that are multiply scattered along their propagation, there is a direct correspondence between the time-of-flight of the ultrasound and the position of the scatterer \cite{Jensen:1996}. Note that, as in any imaging technique, multiple scattering constitutes a strong limitation of the present instrument, especially in the case of concentrated suspensions, for which a specific analysis of the multiply scattered sound should be undertaken \cite{Page:2000}. In the opposite case of a sample that is transparent to ultrasound, seeding the material with acoustic contrast agents allows for a good control of the scattered signal.

In any case the signal recorded on a given reception channel is the sum of the echoes from scatterers at various locations in the imaging plane. An example of such a raw ultrasonic ``speckle'' signal recorded in a dilute aqueous suspension of polydisperse hollow glass spheres (Potters Sphericel, mean diameter 6~$\mu$m, mean density 1.1~g.cm$^{-3}$) is shown as a function of both space and time in Fig.~\ref{f.bscan}(a). Here, the time $t$ corresponds to the time at which the pressure signal is sampled by the receiver, with the origin $t=0$ being the time at which the incident pulse is sent. In other words $t$ is the ultrasonic time-of-flight from and back to the transducer array. The vertical coordinate $z$ denotes the position along the transducer array, the origin being taken at the upper end of the array, which is about 6~mm from the top of the Couette cell [see Fig.~\ref{f.setup}(a)]. This two-dimensional representation of the radio-frequency (RF) backscattered signals $s(t,z)$, also referred to as a ``B-scan'' in the ultrasound literature, clearly shows the echoes of the incident plane wave on the outer cup and on the inner bob as vertical lines at $t\simeq 38.7$ and 41.5~$\mu$s respectively. In between the bob and the cup, the RF signal is constituted of a superposition of spherical wavefronts that correspond to the backscattered echoes from the glass spheres within the gap. 

\paragraph{Procedure for removing fixed echoes. -} A close investigation of a series of $N$ successive RF signals $\{s_i(t,z)\}_{i=1\dots N}$ corresponding to $N$ successive incident plane waves reveals that the spherical wavefronts from the glass spheres, which move along with the scatterers as they are carried away by the shear flow, are mixed with other wavefronts that remain fixed under shear (see supplementary movie 1 with $N=200$)~\cite{Remark:moviesRSI}. These fixed echoes are due to waves diffracted by the edges of the transducer array which get reflected on the various surfaces surrounding the Couette cell (free surface, bottom of the water tank, Couette cell, etc.). Most of these reflections occur before the main plane wave enters the gap of the Couette cell ($t\simeq 38.7~\mu$s) but the corresponding cylindrical echoes may extend over the whole temporal window of interest, as seen in Fig.~\ref{f.bscan}(a) (see also supplementary movie 1)~\cite{Remark:moviesRSI}. Moreover the roughness of the outer cup as well as small defects in the inner bob may also generate fixed spurious echoes.

\begin{figure}
\includegraphics[width=7.5cm]{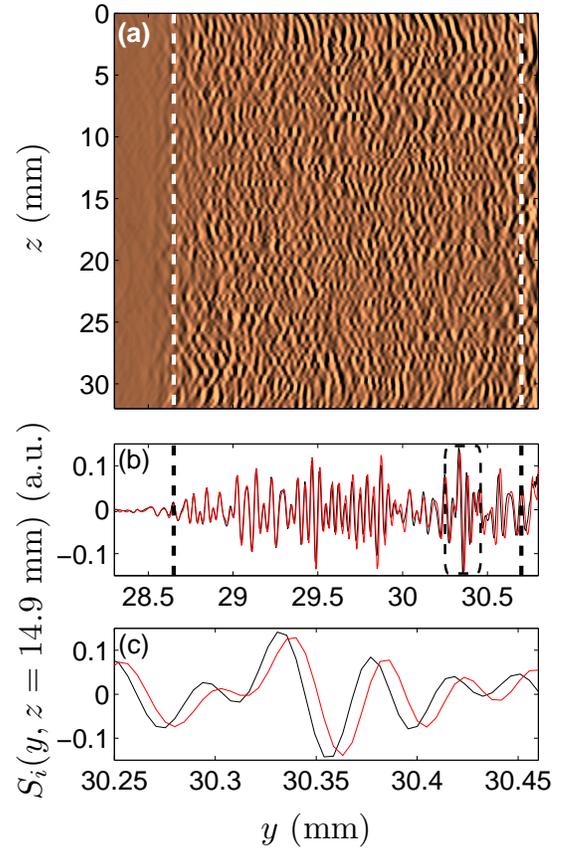}
\caption{\label{f.image}  (a) Beam-formed image $S_i(y,z)$ computed from the corrected speckle signal of Fig.~\ref{f.bscan}(b) and shown as a function of $y$, the distance to the transducer array, and $z$, the position along the array. $S_i$ is normalized by its maximum value and coded in linear color levels. (b) Two successive beam-formed speckle signals $S_i(y,z)$ (in black) and $S_{i+1}(y,z)$ (in red) for a given position $z=14.9$~mm along the transducer array. $S_i$ and $S_{i+1}$ correspond to two different plane wave emissions separated by $\delta t=2$~ms. (c) Enlargement of $S_i(y,z)$ (in black) and $S_{i+1}(y,z)$ (in red) over a window of width $\Delta y=2\lambda$ close to the inner bob [indicated as a dashed box in (b)]. This evidences a noticeable displacement of the speckle to the right when going from $S_i$ to $S_{i+1}$. The dashed lines at $y=28.7$ and 30.7~mm indicate the limits of the gap as inferred from the calibration procedure described in Sect.~\ref{s.calib}. Same experiment as in Fig.~\ref{f.bscan} (see also supplementary movie 2)~\cite{Remark:moviesRSI}.}
\end{figure}

\begin{figure*}
\includegraphics[width=13cm]{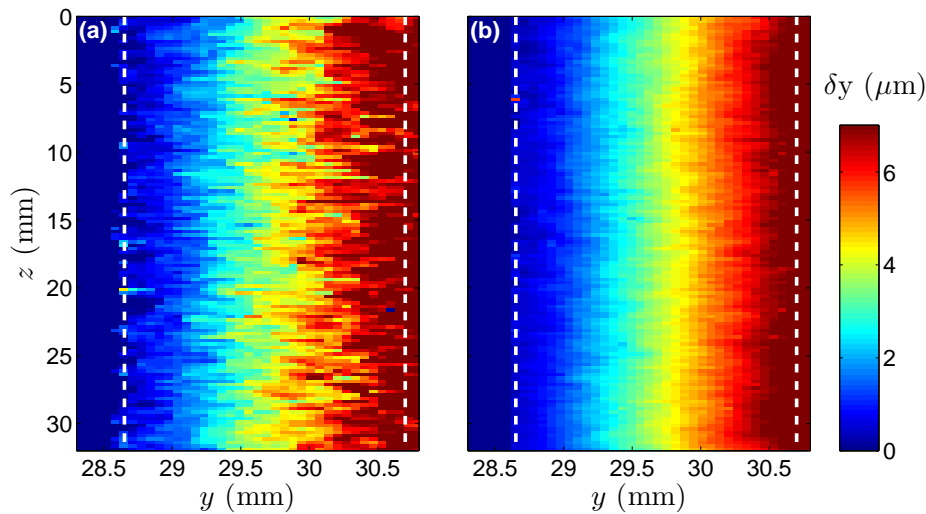}
\caption{\label{f.dpl} (a) Displacement map $\delta y_i(y,z)$ computed from two successive beam-formed images $S_i$ and $S_{i+1}$ separated by $\delta t=2$~ms. (b) Displacement map $\langle\delta y_i(y,z)\rangle_i$ averaged over 199 correlations between 200 successive images. The dashed lines at $y=28.7$ and 30.7~mm indicate the limits of the gap as inferred from the calibration procedure described in Sect.~\ref{s.calib}. Same experiment as in Fig.~\ref{f.bscan}.}
\end{figure*}

As in biomedical imaging where fixed echoes due to artery walls may lead to large artifacts in blood speed detection \cite{Jensen:1996,Ferrara:1996}, we get rid of these spurious echoes by subtracting the speckle signal $\langle s_i(t,z)\rangle_i$ averaged over $N$ successive signals to each individual signal $s_i(t,z)$:
\begin{equation}
\tilde{s}_i(t,z)=s_i(t,z)-\langle s_i(t,z)\rangle_i\,.
\end{equation}
The result obtained from Fig.~\ref{f.bscan}(a) with a series of $N=200$ pulses separated by $\delta t=2$~ms (i.e. a pulse repetition frequency $f_{\rm PRF}=500$~Hz) is shown in Fig.~\ref{f.bscan}(b). In this case, the Newtonian suspension is sheared at a constant shear rate $\dot\gamma=10$~s$^{-1}$. Shear is started long enough before collecting the ultrasonic data so that the flow is laminar and stationary. Figure~\ref{f.bscan}(b) shows that our procedure for removing fixed echoes is very efficient. It remains so as long as all scatterers across the gap move by a significant quantity over the total duration of the pulse sequence so that the (moving) backscattered wavefronts cancel out in the average. In some cases where the fluid may remain fixed in parts of the gap (e.g. in yield stress fluids that show shear localization), the average is taken on a different series of incident pulses performed under a high shear rate, so that the whole sample flows. Since the USV algorithm has already been described as a tool for flow visualization in other contexts \cite{Sandrin:2001,Manneville:2001b}, we briefly recall the main steps of image formation and processing in the following and emphasize only the aspects specific to the case of cylindrical Couette flow imaging.

\paragraph{Computation of beam-formed images. -} In this data processing step based on wave propagation in a single scattering regime, an ultrasonic image $S_i(y,z)$ is formed from a corrected speckle signal $\tilde{s}_i(t,z)$ by using standard parallel beam forming \cite{Sandrin:1999}:
\begin{equation}
S_i(y,z)=\sum_{z_0}\tilde{s}_i(t(y,z,z_0),z)\,,
\label{e.bf}
\end{equation}
where $y$ is the distance from the transducer array along the ultrasonic propagation axis (or ``axial distance'') and
\begin{equation}
t(y,z,z_0)=\frac{y+\sqrt{y^2+(z-z_0)^2}}{c}
\label{e.tof}
\end{equation}
is the ultrasonic time-of-flight from a scatterer located at $(y,z)$ in the gap to the transducer located at height $z_0$, $c$ denoting the sound speed in water. Note that Eq.~(\ref{e.tof}) neither accounts for the sound speed in the outer cup which is made out of PMMA ($c_{\rm PMMA}\simeq 2500$~m.s$^{-1}$) nor from the fact that the sound speed in the sample may differ from that of water. In any case the actual distance $r$ in the sample to the inner bob will be inferred from $y$ through a calibration step described in Sect.~\ref{s.calib}. Moreover the summation in Eq.~(\ref{e.bf}) is taken over 30 values of $z_0$ around $z$. This fastens the data processing without significantly degrading the image resolution. It can also be noted that backscattered echoes are recorded beyond the position of the inner bob (i.e. for $t>41.5~\mu$s in Fig.~\ref{f.bscan} and for $y>30.7$~mm in Fig.~\ref{f.image}). These echoes correspond to the scattering of the reflection of the incident plane wave on the bob \cite{Manneville:2004a}. Therefore only points with $y\lesssim30.8$~mm are considered in Eq.~(\ref{e.tof}).  The beam-formed image computed from Fig.~\ref{f.bscan}(b) is shown in Fig.~\ref{f.image}(a) and the line at $z=14.9$~mm of two successive individual images $S_i(y,z)$ and $S_{i+1}(y,z)$ are shown in Fig.~\ref{f.image}(b) and (c) (see also supplementary movie 2 for a series of $N=200$ successive images separated by 2~ms in a sheared suspension of hollow glass spheres)~\cite{Remark:moviesRSI}. 

\subsubsection{Displacement field}
\label{s.flow}

The speckle displacement along the $y$ direction at point $(y,z)$ between two successive pulses $i$ and $i+1$ is computed by looking for the maximum of the following correlation coefficient as a function of $\delta y$ \cite{Sandrin:2001,Manneville:2004a}:
\begin{equation}
C_i(y,z,\delta y)=\sum_{y'=y-\Delta y/2}^{y+\Delta y/2} S_i(y',z)\,S_{i+1}(y'+\delta y,z)\,,
\label{e.corrcoeff}
\end{equation}
where $\Delta y=2\lambda\simeq 200~\mu$m. An example of correlation window of width $\Delta y$ is shown in Fig.~\ref{f.image}(b) and (c). For a given $(y,z)$, the value of $\delta y$ that maximizes $C_i(y,z,\delta y)$  is computed from a second-order polynomial interpolation of $C_i$ and yields the axial displacement $\delta y_i$ of the speckle between pulses $i$ and $i+1$. In practice, to avoid any redundancy in the velocity data, such a displacement $\delta y_i(y,z)$ is computed only for locations within the gap such that $y_k=y_0+k\Delta y/3$, where $y_0$ corresponds to the beginning of the gap on the cup side and $k$ is an integer. This yields about 30 measurement points across the gap of the Couette cell separated by roughly 65~$\mu$m. A displacement map computed from two successive beam-formed images $S_i$ and $S_{i+1}$ is shown in Fig.~\ref{f.dpl}(a). The velocity gradient is clearly visible even if the level of noise is rather large because only two successive pulses are considered.

Figure~\ref{f.dpl}(b) presents the displacement map obtained by averaging $\delta y_i(y,z)$ over the $N-1=199$ correlations corresponding to the sequence of 200 pulses shown in supplementary movies 1 and 2~\cite{Remark:moviesRSI}. This 0.4~s time-average results in a very small level of noise with a typical variation of less than 5~\% from one channel to another (see also inset of Fig.~\ref{f.calib}). As expected the flow is laminar and homogeneous over the whole height of the Couette cell.

\begin{figure}
\includegraphics[width=9.5cm]{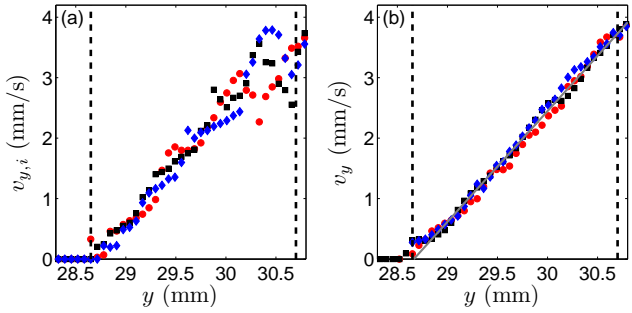}
\caption{\label{f.vy} (a) Axial velocity profiles $v_{y,i}(y,z)$ computed from the two successive beam-formed images $S_i$ and $S_{i+1}$ used in Fig.~\ref{f.dpl}(a) and shown for $z=7.4$ ({\color{blue}$\blacklozenge$}), 15.1 ({\color{red}${\bullet}$}), and 22.4~mm ($\blacksquare$). (b) Axial velocity profiles $v_y(y,z)=\langle v_i(y,z)\rangle_i$ averaged over 199 correlations between 200 successive images. The grey line shows the best linear fit of the full data set averaged over $z$. The dashed lines at $y=28.7$ and 30.7~mm indicate the limits of the gap as inferred from the calibration procedure described in Sect.~\ref{s.calib}. Same experiment as in Fig.~\ref{f.bscan}.}
\end{figure}

Axial velocity maps  are easily deduced from the displacement field by $v_{y,i}(y,z)=\delta y_i(y,z)/\delta t$, where $\delta t=1/f_{\rm PRF}$ is the time interval between two successive pulses. By ``axial velocity'' $v_y$ we mean the projection of the local velocity vector $\mathbf{v}=(v_r,v_\theta,v_z)$ on the ultrasonic propagation axis $y$ [see Fig.~\ref{f.setup}(b)]. This obviously assumes that the scatterers follow the fluid velocity field as passive tracers. Three individual axial velocity profiles $v_{y,i}(y,z)$ deduced from Fig.~\ref{f.dpl}(a) are gathered in Fig.~\ref{f.vy}(a) for three different heights along the transducer array, while Fig.~\ref{f.vy}(b) shows the corresponding time-averages $v_y(y,z)=\langle v_i(y,z)\rangle_i$. 

\section{Calibration and test in a Newtonian fluid}
\label{s.newtonian}

\subsection{Calibration procedure}
\label{s.calib}

In order to recover the velocity field in sheared complex fluids, it is necessary to calibrate the axial velocity data $v_y$. Such a calibration procedure has already been described in detail in Ref.~\cite{Manneville:2004a} and allows one to infer the precise value of the incidence angle $\phi$ and the exact position $y_0$ of the cup wall. In brief, the calibration procedure must be performed in a Newtonian fluid at a low enough shear rate to ensure that the flow is laminar and purely tangential so that $\mathbf{v}=(0,v_\theta,0)$ and $v_y$ simply measures the projection of the orthoradial velocity component $v_\theta$. In our geometry and for a dilute aqueous suspension, this corresponds to $\dot\gamma\lesssim 40$~s$^{-1}$. Linear fits of the $v_y$ vs $y$ data averaged over $N\simeq 200$ pulses and over the vertical direction $z$ for different values of $\dot\gamma$ yield the position $y_0$ of the cup wall with an uncertainty of about 20~$\mu$m [see the fit in Fig.~\ref{f.vy}(b) for $\dot\gamma=10$~s$^{-1}$]. Moreover the angle of incidence $\phi$ is estimated within $\pm 0.1^\circ$ by looking for the value of $\phi$ that best matches the expected velocity profiles for various applied shear rates \cite{Manneville:2004a}.

\begin{figure}
\includegraphics[width=6.5cm]{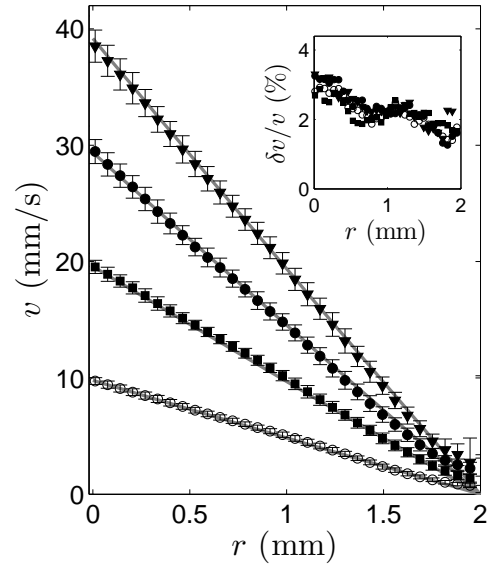}
\caption{\label{f.calib} Tangential velocity profiles $v(r)$ deduced from the calibration procedure [Eq.~(\ref{e.r}) and (\ref{e.v})] with $y_0=28.66$~mm and $\phi=5.2^\circ$ for different applied shear rates: $\dot\gamma=5$ ($\bullet$), 10 ($\square$), 15 ($\bullet$), and 20~s$^{-1}$ ($\blacktriangledown$). Data are averaged over 199 successive correlations and over the vertical direction $z$. Error bars show the standard deviation over $z$. Grey lines are the theoretical predictions for a Newtonian fluid [Eq.~(\ref{e.newt})]. Inset: relative deviation $\delta v/v$ computed as the standard deviation of $v(r,z)$ over $z$ relative to its mean value $v(r)$. Experiments performed in a Newtonian suspension of hollow glass spheres at 1~wt.~\% in water.}
\end{figure}

Once $y_0$ and $\phi$ are calibrated, the axial velocity $v_y(y,z)$ is easily converted to a an {\it apparent} tangential velocity $v(r,z)$, where $r$ denotes the distance to the rotating inner bob, through the relationships:
\begin{eqnarray}
r = \sqrt{R_2^2 + (y-y_0)^2 - 2 R_2 (y-y_0)\cos\phi} - R_1\,,\label{e.r}\\
v(r,z) = \frac{R_1+r}{R_2\sin\phi}\, v_y(y,z)\,.\label{e.v}
\end{eqnarray}
In the small-gap approximation $e/R_1\ll 1$, the previous equations reduce to:
\begin{eqnarray}
r \simeq e - (y-y_0)\sin\phi\,,\label{e.rapprox}\\
v(r,z) \simeq \frac{v_y(y,z)}{\sin\phi}\,. \label{e.vapprox}
\end{eqnarray}
Once again it is crucial to emphasize that $v(r,z)$ deduced from Eqs.~(\ref{e.v}) and (\ref{e.vapprox}) indeed corresponds to the orthoradial component $v_\theta$ of the velocity field $\mathbf{v}$ only when the flow is purely tangential. In the general case of a three-dimensional flow with non-zero radial and vertical components $v_r$ and $v_z$, the USV measurement is $v_y\simeq v_\theta\sin\phi  + v_r \cos\phi$ in the small-gap approximation. Therefore the velocity $v$ defined by Eqs.~(\ref{e.v}) and (\ref{e.vapprox}) is insensitive to the vertical component but may contain a significant contribution from the radial component:
\begin{equation}
v(r,z) \simeq v_\theta(r,z)+ \frac{v_r(r,z)}{\tan\phi}\,.
\label{e.v3D}
\end{equation}

\begin{figure*}
\includegraphics[width=13cm]{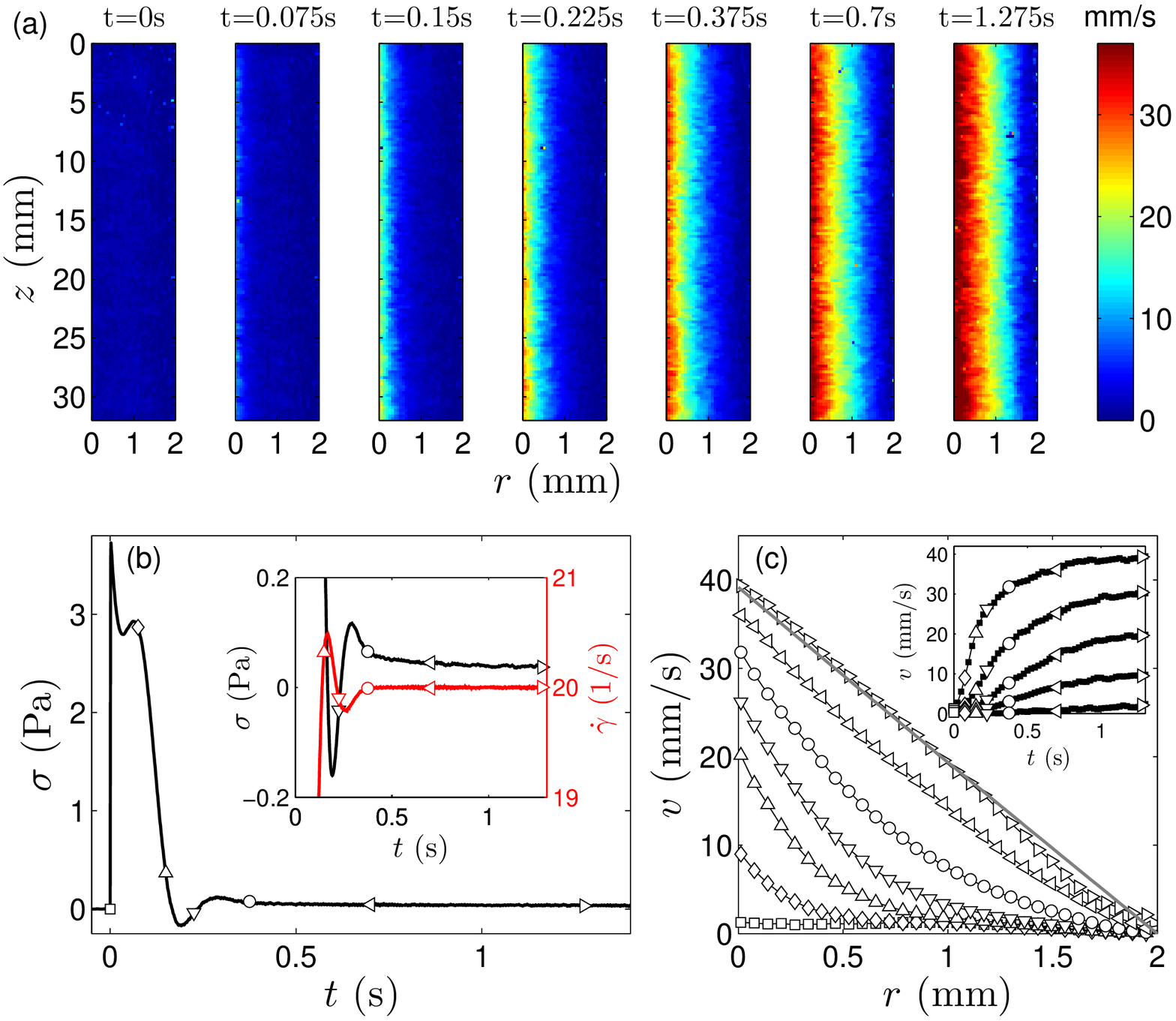}
\caption{\label{f.laminar} Start-up of shear in a Newtonian suspension of hollow glass spheres at 1~wt.~\% in water and sheared at $\dot\gamma=20$~s$^{-1}$ in the laminar regime. (a) Velocity maps $v(r,z,t)$ at different times $t$ indicated on the top row. Each map corresponds to an average over 50 pulses sent every millisecond (see also supplementary movie~3)~\cite{Remark:moviesRSI}. (b) Stress response $\sigma(t)$ recorded simultaneously to the velocity maps. The symbols indicate the times corresponding to the images shown in (a). Inset: magnification of the stress response $\sigma(t)$ (in black) together with the instantaneous shear rate  $\dot\gamma(t)$ (in red) imposed by the rheometer. (c) Velocity profiles $\langle v(r,z,t)\rangle_z$ averaged over the whole height of the transducer array and shown at $t=0$ ($\square$), 0.075 ($\diamond$), 0.15 ($\triangle$), 0.225 ($\triangledown$), 0.375 ($\circ$), 0.7 ($\triangleleft$), and 1.275~s ($\triangleright$) consistently with (a) and (b). The grey line shows the velocity profile expected for a Newtonian fluid in the laminar regime [Eq.~(\ref{e.newt})]. Inset: time evolution of $\langle v(r,z,t)\rangle_z$ at $r=0.01$, 0.46, 0.98, 1.43, and 1.95~mm from top to bottom.}
\end{figure*}

Figure \ref{f.calib} presents the result of the calibration procedure in a suspension of hollow glass spheres at 1~wt.~\% in water. The ultrasonic data are acquired using a single sequence of $N=200$ pulses with a pulse repetition frequency that is tuned depending on the applied shear rate $\dot\gamma$ according to $f_{\rm PRF}=50\dot\gamma$. Whatever the applied shear rate, the velocity profiles almost perfectly coincide with the theoretical predictions for a Newtonian fluid:
\begin{equation}
v(r)=v_0 \left(1+\frac{r}{R_1}\right)\left[ 
\frac{ \left( \frac{R_2}{R_1+r} \right)^2 -1}{\left( \frac{R_2}{R_1} \right)^2 -1}\right]\simeq v_0 \left(1-\frac{r}{e}\right)\,,
\label{e.newt}
\end{equation}
where $v_0$ is the velocity of the inner bob imposed by the rheometer. The last term in Eq.~(\ref{e.newt}) results from the small-gap approximation for which $v_0\simeq\dot\gamma e$. The only significant deviations from the predictions are observed close to the outer cup ($r\simeq 2$~mm) where the velocity is slightly overestimated. Such an artifact can be attributed to our procedure for removing fixed echoes which prevents us from measuring vanishingly small velocities close to the cup.

\subsection{Shear start-up in a Newtonian fluid}
\label{s.start}

In order to highlight the time-resolved capabilities of our rheo-ultrasonic technique, we turn to measurements performed during flow build-up in a Newtonian suspension after shearing at $\dot\gamma=20$~s$^{-1}$ is started at time $t=0$. In the remainder of this paper $t$ shall denote the time after start up of shear (rather than the ultrasonic time-of-flight as in Sect.~\ref{s.setup}). Here one sequence of $N=8,000$ pulses is used with $f_{\rm PRF}=1$~kHz. Figure~\ref{f.laminar}(a) shows a few velocity maps corresponding to averages over 50~successive pulses every 25~ms (see also supplementary movie~3)~\cite{Remark:moviesRSI}. The shear rate and shear stress signals recorded simultaneously by the rheometer are also shown in Fig.~\ref{f.laminar}(b). Since the flow clearly remains homogeneous along the vertical direction during the whole transient, we average the velocity data along $z$ and plot a few velocity profiles and time series $\langle v(r,z,t)\rangle_z$ in Fig.~\ref{f.laminar}(c). 

As expected, the velocity profile is fully developed only for $t\gtrsim\tau$, where $\tau\equiv e^2/\nu \simeq 1$~s is the viscous dissipation time on the size of the entire gap. At earlier times, a boundary layer develops at the rotating inner bob and propagates towards the fixed outer cup. Note that the velocity time series $v(r,z,t)$ in Fig.~\ref{f.laminar}(c) resemble but do not follow exactly the simple error function solution for a boundary layer on an infinite flat plate started impulsively~\cite{Batchelor2000,Kouitat:1990}. The existence of the outer bob, the curvature and closure of the streamlines, and the complex input function for the velocity of the bob all have a non-negligible impact on the transient flow. The complexity of the input function $\dot\gamma(t)$ in the inset of Fig.~\ref{f.laminar}(b) is due to the fact that we are using an intrinsically stress-imposed rheometer working in strain-imposed mode through a feedback loop. In particular the fact that the shear stress presents a strong undershoot into negative values can be attributed to the coupling between the rheometer feedback on the applied torque, the strong inertia of our home-made geometry, and the fluid response. Experiments performed with a smaller, lighter bob would show a faster convergence of the applied shear rate to its target value with much smaller oscillations. Working with a strain-imposed rheometer would allow a better access to the actual fluid response to a step-like shear rate function.

\subsection{Taylor-Couette instability in a Newtonian fluid}
\label{s.tc}

\begin{figure*}
\includegraphics[width=13cm]{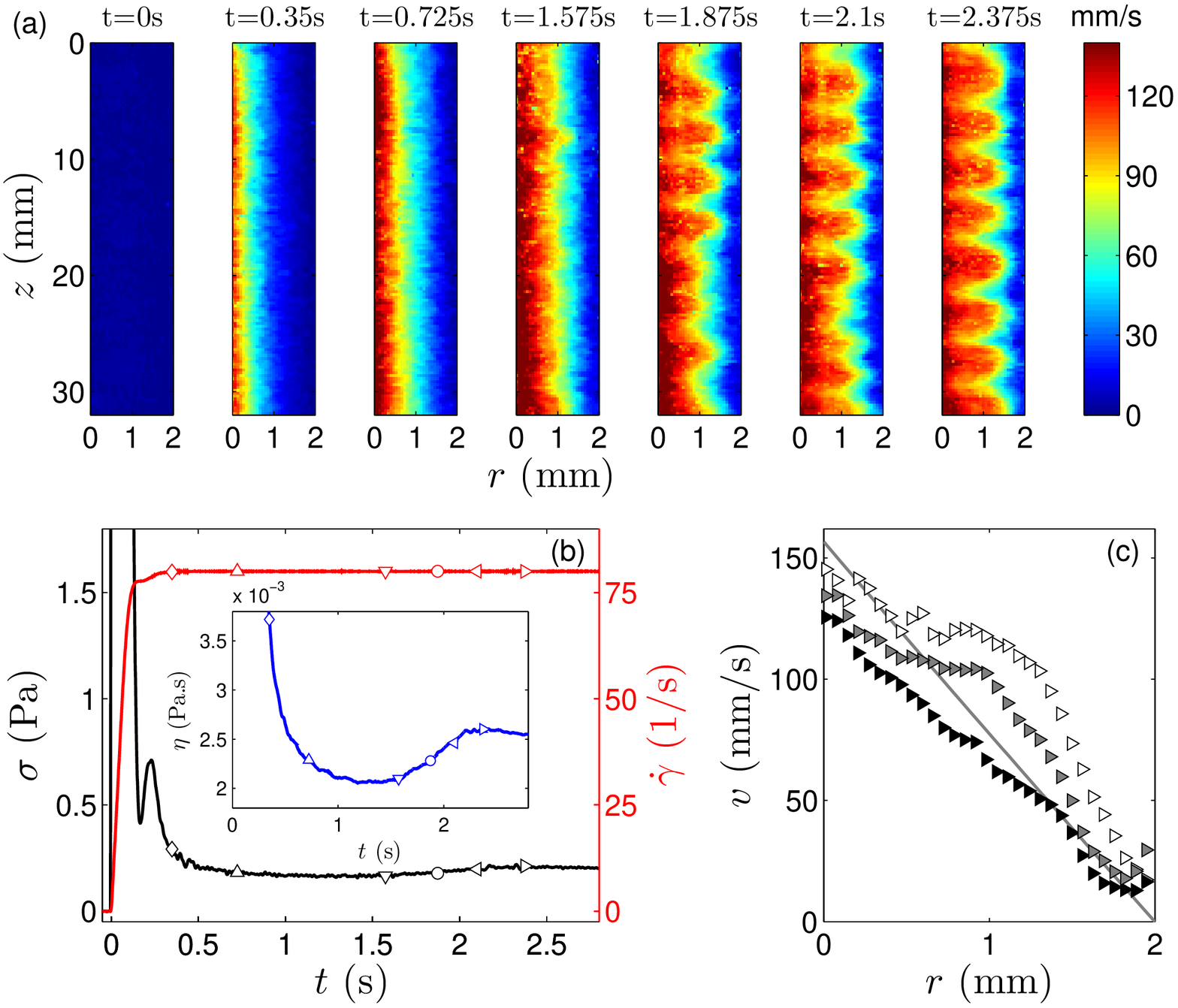}
\caption{\label{f.unstable} Start-up of shear in a Newtonian suspension of hollow glass spheres at 1~wt.~\% in water and sheared at $\dot\gamma=80$~s$^{-1}$ in a vortex flow regime. (a) Velocity maps $v(r,z,t)$ at different times $t$ indicated on the top row. Each map corresponds to an average over 50 pulses sent every 0.5~ms (see also supplementary movie~4)~\cite{Remark:moviesRSI}. (b) Stress response $\sigma(t)$ (in black) recorded simultaneously to the velocity maps together with the instantaneous shear rate  $\dot\gamma(t)$ (in red) imposed by the rheometer. The symbols indicate the times corresponding to the images shown in (a). Inset: apparent viscosity $\eta(t)=\sigma(t)/\dot\gamma(t)$. (c) Velocity profiles $v(r,z,t)$ at $t=2.375$~s and for $z=19.25$ (white symbols, outflow boundary), 20.5 (gray symbols, in between outflow and inflow), and 21.5~mm (black symbols, inflow boundary). The gray line shows the velocity profile expected for a Newtonian fluid in the laminar regime [Eq.~(\ref{e.newt})].}
\end{figure*}

We have seen in the previous section that as long as the velocity field remains purely orthoradial, the measurements at different axial positions do not bring any more information than one-dimensional velocimetry: the velocity maps in Fig.~\ref{f.laminar}(a) are essentially invariant along $z$. The situation changes radically at higher imposed shear rates when this invariance is broken, revealing the full potential of our new technique. For a Newtonian fluid sheared in between concentric cylinders Taylor showed that the purely orthoradial base flow becomes unstable and gives way to a cellular pattern in which the fluid travels in helical paths around the cylinders in layers of counter rotating vortices--now known as Taylor vortices~\cite{Taylor:1923}. In the simplest case where only the inner cylinder is rotating and the gap is small, the vortex flow develops for $Ta \gtrsim 41$, where $Ta\equiv \sqrt{e/R_1} Re$ is the Taylor number (textbooks often use $Ta^2$) and $Re\equiv \dot\gamma\tau$ is the Reynolds number. In the geometry used in our experiments, the threshold of the instability therefore corresponds to $\dot\gamma_c\simeq 50$~s$^{-1}$, which we indeed observe. 

Figure~\ref{f.unstable}(a) shows consecutive velocity maps for a start-up of shear at $\dot\gamma=80$~s$^{-1}$ (see also supplementary movie~4)~\cite{Remark:moviesRSI}. The first few instants are still invariant along $z$ and resemble the transient flow described in section~\ref{s.start}. However, for $t\gtrsim 1.5$~s, the symmetry is progressively lost with the emergence of well defined undulations along $z$. The onset of the vortex flow can also be detected on the transient rheological data in Fig.~\ref{f.unstable}(b). Indeed, the transition from the purely orthoradial flow to the vortex flow brings in an additional flow resistance characterized by an increase in the apparent viscosity $\eta(t)=\sigma(t)/\dot\gamma(t)$ recorded by the rheometer, which is evident on the inset of Fig.~\ref{f.unstable}(b) for $1.5\lesssim t\lesssim 2.5$~s. 

If the flow is observed on longer time scales, the vortex flow is observed to be time periodic as can be checked in supplementary movie~5~\cite{Remark:moviesRSI}. This behavior corresponds to the ``wavy vortex flow'' regime, the outcome of the second instability of the Taylor-Couette system when only the inner cylinder is rotating~\cite{Andereck:1986}. For shear rates closer to the threshold $\dot\gamma_c$, we also observed steady vortex flows (the so-called ``Taylor vortex flow'' regime, for $\dot\gamma_c<\dot\gamma \lesssim 66$~s$^{-1}$, as expected~\cite{Andereck:1986}).    

Figure~\ref{f.unstable}(a) shows that the main orthoradial flow is deformed by the influence of the secondary vortex flow~\cite{Davey:1962}. Typically, at locations along $z$ that correspond respectively to the outflow/inflow boundary of vortices (strong outward/inward radial flow), the main flow is pushed outward/inward. Figure~\ref{f.unstable}(c) gives examples of such deformed individual velocity profiles at the locations of an outflow boundary (white symbols), an inflow boundary (black symbols) and in between (gray symbols). Thus, for each wavelength of the undulations on the velocity maps, there is a pair of counter-rotating vortices. The large field of view of our imaging technique allows us to visualize more than 7 wavelengths, yielding a value of $4.1\pm0.1$~mm, in agreement with the theoretical prediction $\lambda_c\simeq 2e=4$~mm computed by Taylor~\cite{Taylor:1923}.

To the best of our knowledge a characterization of the Taylor-Couette instability solely based on a time resolved velocity map of the orthoradial velocity profiles $v_\theta(r,z,t)$ at one single location along $\theta$ does not exist in the literature. Wereley and Lueptow rather used PIV to measure $v_r$ and $v_z$ in a slice of the gap at a given angle $\theta$~\cite{Wereley:1998}. Then from the orthoradial component of the momentum balance they could reconstruct orthoradial velocity maps similar to those measured here. Later, Akonur and Lueptow used PIV on various slices to reconstruct a three dimensional velocity map~\cite{Akonur:2003}. Nonetheless, the measurements at various locations along $z$ were obtained independently at different times and required phase alignment~\cite{Akonur:2003}. A similar procedure was used by Raguin and Georgiadis on magnetic resonance velocimetry data of a Taylor-Couette-Poiseuille flow (i.e. with superimposed axial flow)~\cite{Raguin:2004}. In contrast to these earlier results, our new ultrafast ultrasonic imaging technique provides a way to monitor the flow over a large range of vertical positions simultaneously and with a temporal resolution that is fast enough to capture unsteady flows.  
  
Note that according to Eq.~(\ref{e.v3D}) the data displayed in Fig.~\ref{f.unstable}(a) and (c) are not exactly equal to the orthoradial velocity component. The vertical oscillations on the velocity maps are enhanced by the contribution of the radial velocity to the measurement. Moreover, the velocities measured near the inner and outer cylinders deviate from the expected values more significantly than in the purely orthoradial flow regime used for calibration in Fig.~\ref{f.calib}. Once again, this could be an artefact due to the procedure used to remove fixed echoes, or it could also be a consequence of the secondary flow. Indeed, the vortices create localized jets near the walls that can segregate the tracers~\cite{Akonur:2003}. 

\section{Illustration in a wormlike micellar solution}
\label{s.wlm}

\begin{figure*}
\includegraphics[width=13cm]{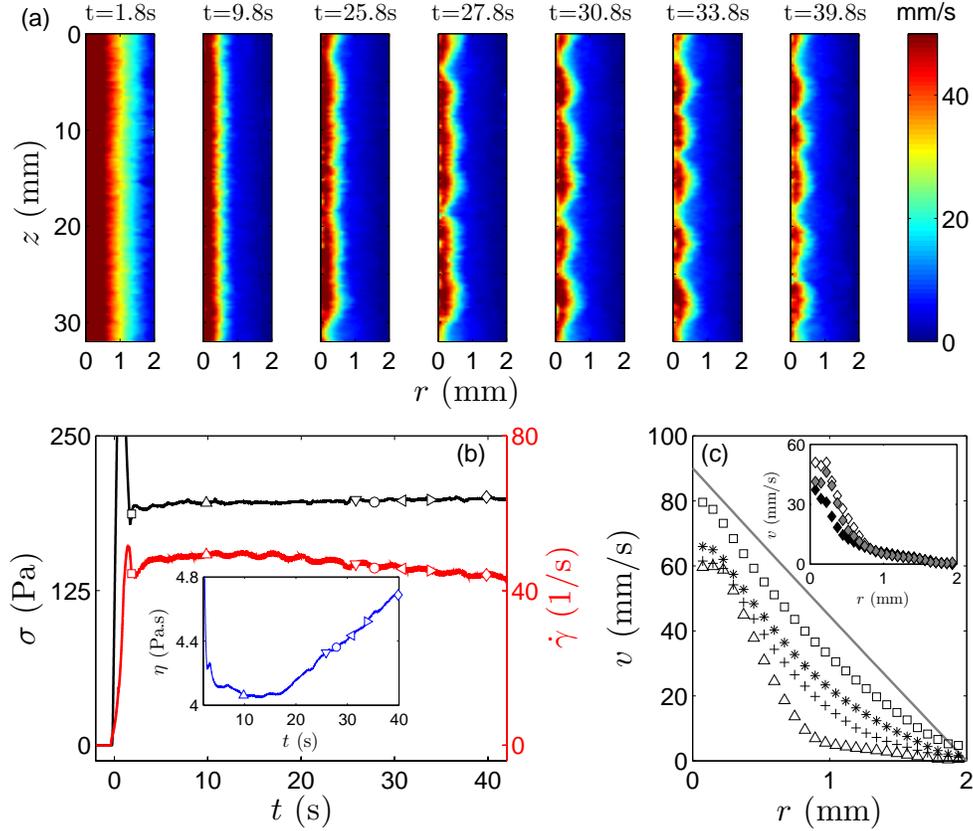}
\caption{\label{f.micelles} Start-up of shear in a solution of wormlike micelles ([CTAB]=0.3 M, [NaNO$_3$]=0.34 M) seeded with hollow glass spheres at 1~wt.~\% and sheared at $\dot\gamma=46\pm 4$~s$^{-1}$ in the shear-banding regime with Taylor-like vortices. (a) Velocity maps $v(r,z,t)$ at different times $t$ indicated on the top row. Each map corresponds to an average over 100 pulses sent every 1~ms (see also supplementary movie~6)~\cite{Remark:moviesRSI}. Each of the $N_{\rm seq}=200$ sequences of $N=100$ pulses is separated from the next one by 0.5~s. (b)~Stress response $\sigma(t)$ (in black) imposed by the rheometer recorded simultaneously to the velocity maps together with the instantaneous shear rate  $\dot\gamma(t)$ (in red). The symbols indicate the times corresponding to the images shown in (a). Inset: apparent viscosity $\eta(t)=\sigma(t)/\dot\gamma(t)$. (c) Velocity profiles $\langle v(r,z,t)\rangle_z$ averaged over the whole height of the transducer array and shown at $t=1.8$ ($\square$), 2.3 ($*$), 2.8 ($+$), and 9.8~s ($\triangle$). The gray line shows the velocity profile expected for a Newtonian fluid in the laminar regime [Eq.~(\ref{e.newt})]. Inset: velocity profiles $v(r,z,t)$ at $t=39.8$~s and for $z=10.5$ (white symbols, outflow boundary), 9.5 (gray symbols, in between outflow and inflow), and 8.5~mm (black symbols, inflow boundary). }
\end{figure*}

To illustrate our new imaging technique on a non-Newtonian flow of soft matter, we choose an aqueous surfactant solution of cetyltrimethylammonium bromide (CTAB) at concentration of 0.3 M and sodium nitrate (NaNO$_3$) at concentration of 0.34 M. The temperature is fixed at $T=28\pm 0.1^\circ$C. In this regime of concentration, salinity and temperature, the surfactant molecules aggregate in long, flexible wormlike micelles that entangle in a way similar to classical polymers. Such solutions of wormlike micelles have been considered as model systems for rheological research for more than twenty years~\cite{Rehage:1991,Berret:2005,Cates:2006,Lerouge:2010}. They have a Maxwellian behavior in the linear rheology regime (i.e. they have a unique viscoelastic relaxation time $\tau_v$), and they exhibit shear-banding and viscoelastic instabilities in the non-linear regime~\cite{Fardin:2012d}. 

For a purely viscoelastic fluid sheared in between concentric cylinders, Larson, Shaqfeh and Muller showed that the purely orthoradial base flow becomes unstable and is replaced by a cellular pattern reminiscent of the Taylor vortices but driven by normal stresses rather than by inertia~\cite{Larson:1990,Larson:1992,Morozov:2007}. Derivations from simple viscoelastic models like the Upper Convected Maxwell model~\cite{Larson:1990} and experiments on polymer solutions showed that in this case the dimensionless number controlling the instability is a viscoelastic Taylor number $\Sigma\equiv \sqrt{e/R_1} Wi$, where $Wi\equiv \dot\gamma\tau_v$ is the Weissenberg number~\cite{Larson:1990}. Recently, it was shown that shear-banding wormlike micelles solutions also exhibit a purely viscoelastic instability~\cite{Fardin:2012d}. The base flow is not homogeneous but shear-banded, which means that the velocity profile in a shear flow presents two distinct slopes defining a band of high shear rate $\dot\gamma_h$ (close to the rotating inner bob) next to a band of low shear rate $\dot\gamma_l$ (close to the fixed outer cup)~\cite{Manneville:2008}. The proportion $\alpha$ of the high shear rate band increases roughly linearly with the global shear rate between $\dot\gamma_l$ and $\dot\gamma_h$~\cite{Lettinga:2009,Salmon:2003c,Hu:2005,Fardin:2012c}. In this case, the instability originates solely in the high shear rate band and Taylor-like vortices develop in this band for a viscoelastic Taylor number rescaled on the size of the high shear rate band $\Sigma^*\equiv \sqrt{\alpha e/R_1} Wi_h$, where $Wi_h\equiv \dot\gamma_h\tau_v$ is the Weissenberg number in the high shear rate band~\cite{Fardin:2012d}. 

Figure~\ref{f.micelles}(a) shows consecutive velocity maps for a start-up of shear at $\dot\gamma=46\pm 4$~s$^{-1}$ (see also supplementary movie~6)~\cite{Remark:moviesRSI}. In order to follow the flow dynamics over about one minute while keeping up with the memory constraints of the ultrasonic scanner, the multiple sequence acquisition procedure described in section~\ref{s.plane} is used with $N_{\rm seq}=200$ sequences of $N=100$ pulses at $f_{\rm PRF}=1$~kHz. Sequences are separated by 0.5~s. Here, the rather poor control of the applied shear rate [to within about 10~\%, see Fig.~\ref{f.micelles}(b)] is due to the coupling between the rheometer feedback loop, the strong sample viscoelasticity, and the large inertia of our home-made Couette geometry. 

During the first few seconds of the transient, the flow remains invariant along $z$ and shows the development of a shear-banded velocity profile, in accordance with the rich literature on the subject~\cite{Lerouge:2010}. More precisely, Fig.~\ref{f.micelles}(c) displays a few velocity profiles $\langle v(r,z,t)\rangle_z$ averaged along $z$ during the nucleation of the shear band for $t\lesssim 10$~s. The evolution of these velocity profiles is fully consistent with previous works~\cite{Hu:2005,Miller:2007,Lettinga:2009}. Note that the establishment of the shear-banded velocity profiles is also associated with an enhanced wall slip at the moving inner cylinder. This fact seems to be independent of the existence of secondary flows, as already reported in~\cite{Lettinga:2009, Fardin:2012a, Fardin:2012b}.  

At later stages (for $t\gtrsim 10$~s), the vortex flow emerges and deforms the interface between the shear bands. In the asymptotic state (for $t\gtrsim 35$~s), for each wavelength of the undulations on the velocity map, there is a pair of counter-rotating vortices, the size of which scales with the width of the high shear rate band~\cite{Fardin:2012d}. As shown in the inset of Fig.~\ref{f.micelles}(b), the emergence of the vortex flow generates a viscosity increase, which is already well documented~\cite{Fardin:2012d} and is reminiscent of that seen in Fig.~\ref{f.unstable}(b). As in the Newtonian case described in section~\ref{s.tc}, the emergence of the vortex flow is monitored through its impact on the main orthoradial flow [see rightmost images in Fig.~\ref{f.micelles}(a)]. In the inset of Fig.~\ref{f.micelles}(c), examples of shear-banded velocity profiles deformed by the secondary vortex flow are shown at the locations for an outflow boundary (white symbols), an inflow boundary (black symbols) and in between (gray symbols). When compared to the Newtonian case shown in Fig.~\ref{f.unstable}(c), individual velocity profiles in wormlike micelles are affected by the secondary flow only in the highly sheared region (for $r\lesssim 1$~mm) since vortices develop only in the high shear band.

\section{Conclusion and perspectives}
\label{s.conclu}
The above results show that our new ultrafast ultrasonic imaging technique is a powerful tool for measuring spatiotemporal variations of the main flow of sheared fluids, simultaneously to traditional rheometry. We have successfully tested the technique on a Newtonian fluid that undergoes a purely inertial Taylor-Couette instability, and on a non-Newtonian fluid that exhibits shear-banding and undergoes a purely elastic Taylor-Couette instability. In both cases, our technique can monitor the emergence of secondary flows through their impact on the main flow, which loses its invariance along the vertical direction $z$. Our excellent time resolution gives access to the transient establishment of secondary flows and to their eventual spatiotemporal dynamics. 

In the past few years the existence of secondary flows has been raised as a possible explanation for the often complicated signals recorded by rheometers on flows of complex fluids, which are usually assumed to be viscometric. In future work, we plan to use our new imaging technique to test for the presence of spatially and temporally heterogeneous flows in a variety of complex fluids, which do not need to be optically transparent but only need to scatter ultrasound in the single scattering regime.
Indeed, our instrument allows us to probe any possible flow heterogeneity along the radial and/or vertical direction. For instance, as shown in the present paper, the flow of wormlike micelles becomes inhomogeneous first along the radial direction $r$ because of the shear-banding instability and subsequently along the vertical direction $z$ due to secondary flows triggered by an elastic instability. It is also possible for some complex fluids to first become heterogeneous along $z$ due to a constitutive instability similar to shear banding and coined ``vorticity banding''~\cite{Olmsted:2008}. On the other hand, yield stress fluids, such as emulsions, gels, muds, clay suspensions, or colloidal pastes, are known to present shear localization along $r$, in which a liquidlike band coexists with a solidlike region under simple shear~\cite{Schall:2010}. In any case, our technique is very well suited to investigate these various materials in the Taylor-Couette geometry. Ultimately we plan on extending our technique to other flows commonly used in rheological research, such as the cone-and-plate and plate-and-plate geometries. 

\begin{acknowledgments}
We acknowledge invaluable technical help from D. Israel at TA Instruments and J. Lach\`evre at Lecoeur Electronique. T. Divoux and S. Lerouge are thanked for fruitful discussions. We acknowledge funding from the European Research Council under the European Union's Seventh Framework Programme (FP7/2007-2013) / ERC grant agreement n$^\circ$~258803. 
\end{acknowledgments}


%

\end{document}